\documentclass{elsart5p}
\usepackage[UKenglish]{babel}
\usepackage{amssymb}
\usepackage{graphicx}
\usepackage{cite}
\usepackage{lineno}

\begin{document}

\begin{frontmatter}

\title{Measuring the proton spectrum in neutron decay -- latest results with $a$SPECT}

\author[ill,e18]{M.~Simson\corauthref{cor}}\corauth[cor]{Corresponding
author. Tel.: +33-47-620-7623; fax: +33-47-620-7777; email: \texttt{simson@ill.fr}.},
\author[mainz]{F.~Ayala Guardia}, \author[virginia]{S.~Bae{\ss}ler}, \author[mainz]{M.~Borg}, \author[karls,bud]{F.~Gl\"{u}ck}, \author[mainz]{W.~Heil}, \author[e18]{I.~Konorov}, \author[mainz]{G.~Konrad}, \author[mainz]{R.~Mu\~noz Horta}, \author[ill,e18]{K.K.H.~Leung}, \author[mainz]{Yu.~Sobolev}, \author[ill]{T.~Soldner}, \author[e18]{H.-F.~Wirth\corauthref{lmu}}\corauth[lmu]{Present address: Ludwig-Maximilians-Universit\"at M\"un\-chen, Fa\-kul\-t\"at f\"ur Physik, 85748 Garching, Germany.}, \author[ill,e18]{O.~Zimmer}

\address[ill]{Institut Laue-Langevin, 6, rue Jules Horowitz, 38042 Grenoble, France}
\address[e18]{Physik-Department E18, TU M\"{u}nchen, 85748 Garching, Germany}
\address[mainz]{Institut f\"{u}r Physik, Universit\"{a}t Mainz, 55099 Mainz, Germany}
\address[virginia]{University of Virginia, Charlottesville, VA 22904, U.S.A.}
\address[karls]{IEKP, Universit\"{a}t Karlsruhe (TH), 76131 Karlsruhe, Germany}
\address[bud]{KFKI, RMKI, Budapest POB 49, Hungary}

\begin{abstract}
The retardation spectrometer $a$SPECT was built to measure the shape of the proton spectrum in free neutron decay with high precision. This allows us to determine the antineutrino electron angular correlation coefficient $a$. We aim for a precision more than one order of magnitude better than the present best value, which is $\Delta a /a = 5\,\%$.

In a recent beam time performed at the Institut Laue-Langevin during April\,/\,May 2008 we reached a statistical accuracy of about $2\,\%$ per 24 hours measurement time. Several systematic effects were investigated experimentally. We expect the total relative uncertainty to be well below 5\,\%.
\end{abstract}

\end{frontmatter}

\sloppy
\section{Introduction}
\label{Introduction}
The decay of the free neutron allows one to determine the coupling constants of the weak interaction and can be used to search for physics beyond the standard model of elementary particle physics \cite{Sev06, Abel08}. The decay rate for unpolarised neutrons can be described as \cite{Jack57}:

\begin{eqnarray}
{\mathrm d}\Gamma & \propto & \left( 1+a\frac{{\vec p_{\rm e}}{\vec p_\nu}}{E_{\rm e}E_\nu}+b\frac{m_{\rm e}}{E_{\rm e}}\right) {\mathrm d}\Omega_{\rm e}{\mathrm d}\Omega_\nu {\mathrm d}E_{\rm e}
\end{eqnarray}
where $m_{\rm e}$ is the mass of the electron and $\vec p_{\rm e}$, $\vec p_\nu$, $E_{\rm e}$ and $E_\nu$ are the momenta and energies of electron and antineutrino, respectively.

Within the standard model, where neutron decay is described as a V--A type interaction, the antineutrino electron angular correlation coefficient $a$ depends only on the ratio of the axial vector and vector coupling constants $\lambda=g_{\rm A}/g_{\rm V}$: 
\begin{equation}
a= \frac{1-\lambda^2}{1+3\lambda^2}.
\label{eq:CorrelationLambdaFirst}
\end{equation}

Since $\lambda$ describes the renormalisation of the axial vector current by the structure of the nucleon, it cannot be calculated well enough from first principles and thus has to be determined experimentally.

From $\lambda$ and the neutron lifetime, the element $V_{\rm ud}$ of the Cabibbo-Kobayashi-Maskawa matrix can be derived. Furthermore, the determination of $\lambda$ from different observables (e.g. from $a$ and the beta asymmetry $A$ \cite{Jack57}) permits cross-checks of the theory and searches for non-V--A couplings \cite{Sev06}.

\section{The spectrometer}
The neutrino is hard to detect, hence we infer $a$ from the shape of the proton recoil spectrum. $a$SPECT is a retardation spectrometer. This means, the spectrum is measured by counting all decay protons that overcome a potential barrier. By varying the height of the barrier the shape of the proton spectrum can be reconstructed.

\begin{figure}[t]
\center
\includegraphics[width = 0.40\textwidth]{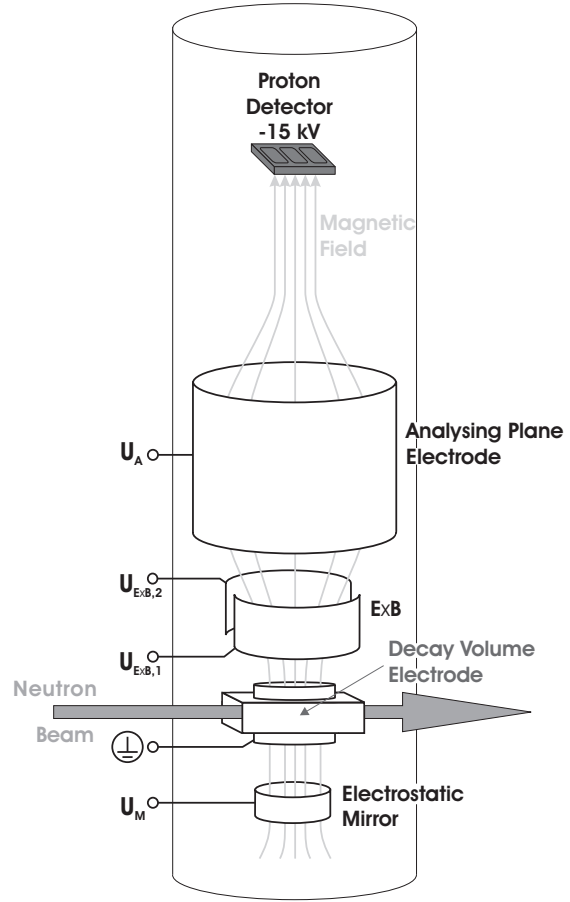}
\caption{Simplified sketch of the $a$SPECT spectrometer (a detailed picture can be found in \cite{bae08}). See text for description.} \label{sketch}
\end{figure}

A beam of cold, unpolarised neutrons passes through the spectrometer, where about $10^{-8}$ of the neutrons decay in the decay volume (see fig.\,\ref{sketch}). This region is held at ground potential, from here the decay protons are guided to the proton detector by a strong magnetic field (about 2\,T in the decay volume during the beam time discussed below). About half of the protons are emitted in the opposite direction. These are reflected by an electrostatic mirror and thus finally all protons are directed towards the detector. The endpoint energy of the proton spectrum is about 751\,eV, therefore a voltage of $U_{\rm M}=820$\,V at the electrostatic mirror reflects all protons. Before the protons can reach the detector, they have to overcome a potential barrier, which is maximal at the so-called analysing plane. The barrier potential is generated by a 54\,cm long cylindrical electrode, held at a voltage $U_{\rm A}$. In the analysing plane the magnetic field is about 5 times lower than in the decay volume. Protons travel from the decay volume to the analysing plane, gyrating about a magnetic field line. As a result of the adiabatic invariance of the magnetic moment, part of the proton's momentum transverse to the field line is transferred into parallel momentum. In the adiabatic approximation, the probability that a proton overcomes the potential barrier can be analytically calculated as a piecewise function:
\begin{equation}
F_{\rm tr}(T) = \left\{
  \begin{array}{ll}
    0 & ;~T < eU_{\rm A} \\
    1-\sqrt{1-\left(1-\frac{eU_{\rm A}}{T}\right)/r_{\rm B}} & ;~\textrm{otherwise} \\
	1 & ;~T > \frac{eU_{\rm A}}{1-r_{\rm B}}
  \end{array} \right. ,
\label{eq:TransmissionFunction}
\end{equation}
where $T$ is the kinetic energy and $e$ the charge of the proton. This transmission function $F_{\rm tr}(T)$ depends only on $U_{\rm A}$ and the ratio of the magnetic fields in the analysing plane and the decay volume, $r_{\rm B} = B_{\rm A}/B_{\rm 0}$.

Further details of the spectrometer design and systematic effects may be found in \cite{Glueck,bae08}.

After the electrostatic barrier the magnetic field increases to focus particles that overcome the barrier onto a silicon drift detector (SDD) \cite{sddreadout}. The detector is held at a high negative potential to post-accelerate the protons to detectable energies, and to ensure that the protons overcome the magnetic mirror created by the increasing field. A SDD is a semiconductor detector based on the principle of sidewards depletion \cite{sdd}, which allows the depletion of a large detector volume with a small readout node. In the spectrometer, one chip with 3 detector pads ($10\times10$\,${\rm mm}^2$ each) is used. The first amplifying transistor as well as a diode for temperature measurement are implemented on the chip.\footnote{Results with a smaller test detector have been published in \cite{sim07}.} The detector signal is amplified and digitised by a continuously sampling 12 bit analogue to digital converter with a 20\,MHz clock. The data is then analysed on-line with respect to a triggering algorithm by a FPGA (Field Programmable Gate Array) which sends a $5\,\mu$s long event to a storage PC in case of a trigger. This event length is the dead time of the detection system.

\begin{figure}[t]
\center
\includegraphics[width = 0.45\textwidth]{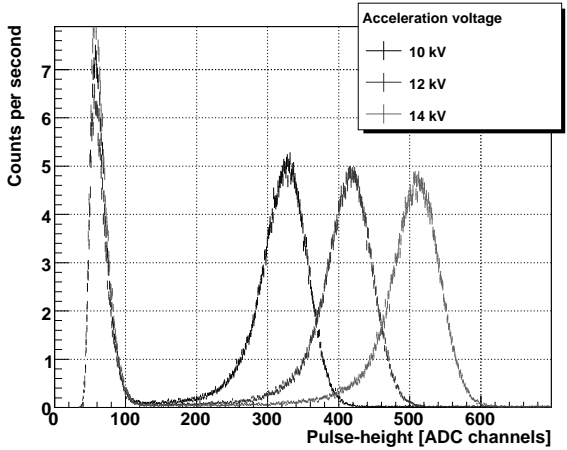}
\caption{Typical pulse-height spectra from the beam time, taken with the silicon drift detector at different acceleration voltages. By increasing the acceleration voltage the proton peak (right peak) is shifted to higher ADC channels, whereas the position of the electronics noise (left peak) is not influenced.} \label{spectra}
\end{figure}

The SDD allowed us to significantly reduce the acceleration voltage, compared to the previously used silicon PIN diode detector. Thus we avoided problems like electrical breakdowns or instabilities of the background due to field emission. During the beam time post-acceleration voltages from $-10$ to $-15$\,kV were used. Some pulse-height spectra with different acceleration voltages are shown in  fig.\,\ref{spectra}.

Protons hit the detector with slightly different incident angles and energies due to the distribution of initial angles and energies, thus an angle or energy dependence in the detection probability leads to a systematic effect. Although first simulations show that this effect is rather small, it will be investigated experimentally at a proton source dedicated for detector tests \cite{mue07}.

$a$SPECT was set up at the cold neutron beam line PF1B \cite{abe2006} at the Institut Laue-Langevin. The neutron beam was shaped by apertures made from sintered boron carbide in the beam tube before the spectrometer (pressure $\sim10^{-2}$\,mbar) and apertures from isotopically enriched $^6$LiF inside the spectrometer ($\sim5\times10^{-9}$\,mbar). The vacua were separated by 250\,$\mu$m thick MgAl3Zn1 windows. A fast neutron shutter in front of the beam tube permitted to open and close the neutron beam during data taking. Behind the spectrometer, a beam stop from boron carbide with an integrated neutron flux monitor was installed. The thermal neutron flux at the exit window of the spectrometer was $6\times 10^9\,{\rm n}/{\rm cm^2s}$. Outside the spectrometer the beam was shielded by several layers of $^{10}$B loaded rubber and 10\,cm of lead.

\section{Investigations of systematic effects} 
All information in this section is preliminary.

The proton spectrum was determined by setting the analysing plane electrode to 7 different voltages $U_{\rm A}$. 50\,V is used for the normalisation of the count rates, 400\,V approximately provides the best statistical sensitivity towards $a$ \cite{aspect2000}. The background was quantified by measurements with 780\,V. Most of the background is caused by beam-related electrons from neutron decay which are also guided towards the detector. At 780\,V no proton can overcome the barrier, whereas the background is only marginally influenced. The pure proton count rate is obtained by subtracting the background from the measurements with lower voltages. To learn about systematic effects and to gain a more precise knowledge of the shape of the spectrum 0\,V, 50\,V, 250\,V, 400\,V, 500\,V, 600\,V, and 780\,V were measured.

Typically, about 470 events per second with pulse-heights in the integration window from 160 to 900 ADC channels (fig.\,\ref{spectra}) at 50\,V analysing plane voltage were counted on one detector pad. At $U_{\rm A}=780$\,V  the count rate in the same window was about 7 counts per second. This means, approximately 460 protons per second were detected. With closed neutron shutter the count rate dropped to 0.2 counts per second in the integration window.

In our previous beam time, the background count rate without neutron beam at $U_{\rm A}=780$\,V was higher than at $U_{\rm A}=50$\,V by several Hz \cite{bae08}. For this beam time, changes were made to the detector high voltage and the rest of the electrodes system to suppress the dependence of the background count rate on the analysing plane voltage. So far, such a dependence was not clearly identified in the present beam time. For example, for a run of 16 hours, the background count rate measured during the periods with closed neutron shutter was $0.197(18)$\,Hz for $U_{\rm A}=50$\,V and $0.154(20)$\,Hz for $U_{\rm A}=780$\,V.

The absolute height of the potential barrier has to be known precisely to determine the transmission function, eq.\,(\ref{eq:TransmissionFunction}). The voltage applied to the electrode is monitored by a precise multimeter\footnote{Agilent 3458A.}. The accuracy of the voltage settings was better than 5\,mV, limited by the multimeter's calibration. Stability and reproducibility of the voltage were much better. However, the electrostatic potential inside the electrode is affected by the work function of the electrode surface. In a cylindrical sample electrode at room temperature, a variation of up to 100\,mV was found. To investigate this effect, further measurements with a Kelvin probe are ongoing.

Protons and positive ions with too low energy to overcome the potential barrier are trapped between the electrostatic mirror and the analysing plane. They are removed by an $\vec{E}\times\vec{B}$ electrode between the decay volume and the analysing plane (see fig.\,\ref{sketch}). This electrode consists of two half cylinders set to different voltages $U_{\rm E\times B,1}$ and $U_{\rm E\times B,2}$. Trapped particles oscillate between the mirror and the analysing plane. Every time they pass the electrical field produced by the $\vec{E}\times\vec{B}$ electrode they drift into the same direction, perpendicular to both $\vec{E}$ and $\vec{B}$. 

In addition, electrons and negative ions may be trapped in the region of the analysing plane electrode.

\begin{figure}[t]
\center
\includegraphics[width = 0.45\textwidth]{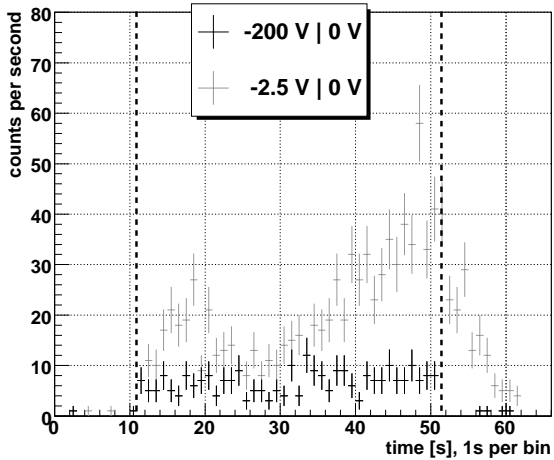}
\caption{The background count rate in the integration window for two different settings of the $\vec{E}\times\vec{B}$ electrode. The dashed lines show where the neutron shutter was opened (after $\sim11$\,s) and later closed again ($\sim52$\,s).} \label{ExB}
\end{figure}

These trapping effects were investigated with two methods:
\begin{enumerate}
\item For each analysing plane voltage $U_{\rm A}$, the measurement started with closed neutron shutter during a time $t_1$, then the shutter was opened for $t_2$, and closed again for $t_3$. This mode allowed us to investigate the environmental, not beam-related background and possible differences in the background count rate before and after the shutter was opened. A higher count rate in the last part would point to either trapped particles inside the spectrometer or activation of some material. Typically, the total measurement time with closed shutter was about half as long as the time with open shutter ($t_1+t_3 \approx \frac{1}{2}t_2$). 
\item The analysing plane was set to 780\,V so that no protons should be able to overcome the barrier and the $\vec{E}\times\vec{B}$ drift potential was reduced from the standard setting of $U_{\rm E\times B,1}\left|\right.U_{\rm E\times B,2}$ = $-1000$\,V\,$\left|\right.-50$\,V on the two half cylinders of the electrode in several steps down to $-2.5$\,V\,$\left|\right.0$\,V. A comparison of two measurements with $-2.5$\,V\,$\left|\right.0$\,V and $-200$\,V\,$\left|\right.0$\,V is shown in fig.\,\ref{ExB}. For the lowest drift voltage of $2.5$\,V there are only some counts before the shutter is opened, this is the environmental background. As soon as the shutter is opened, protons are trapped and the count rate starts to fluctuate. Even after the shutter is closed again, there are still particles that reach the detector. Already a potential of $-10$\,V\,$\left|\right.0$\,V reduces this behaviour strongly, and with $-200$\,V\,$\left|\right.0$\,V the count rate is stable as long as the shutter is open and drops back to the environmental background count rate immediately after the shutter is closed again.
\end{enumerate}

The knowledge of the magnetic field in both decay volume and analysing plane is important for the calculation of the transmission function, eq.\,(\ref{eq:TransmissionFunction}). To determine the magnetic field ratio $r_{\mathrm B}$, the magnetic field has been measured in the spectrometer installed at the beam position, before and after the beam time. There was no significant change in the ratio or the absolute values of the fields. To check the response of the spectrometer, the ratio was changed by about $1\,\%$ by additional, external coils. These coils do not change the homogeneity or the shape of the magnetic field in both analysing plane and decay volume. As expected, the count rates for the different measured voltages $U_{\rm A}$ changed due to the altered transmission function, but the value for $a$ is, within statistical errors, still the same if we use the adapted ratio of the magnetic fields $r_{\mathrm B}$ in eq.\,(\ref{eq:TransmissionFunction}).

To monitor the ratio $r_{\mathrm B}$, it is planned to install a nuclear magnetic resonance system with polarised $^3$He. The NMR frequency will be measured in two cells close to decay volume and analysing plane, respectively. The ratio of those frequencies is proportional to $r_{\mathrm B}$.

The superconducting magnet coils show some hysteresis which might alter the field values if the magnet is set to a different strength. Measurements showed that despite the hysteresis magnetic field settings can be reproduced if for each change in the coil setting the same procedure is used, which starts with warming up the coils above their critical temperature. Then they are cooled down again and finally ramped directly to the desired current.

Another systematic effect we investigated is the so-called edge effect. The neutron beam profile in the decay volume is projected onto the detector by the magnetic field. Close to the edge of the detector, two cases may occur: On the one hand, protons produced outside the directly projected volume can hit the detector, since they gyrate about a magnetic field line. This increases the detected count rate. On the other hand, protons produced within the projected area might miss the detector, resulting in a loss of detected count rate. As the gyration radius of a proton depends on its momentum, the proton count rate gain and loss is momentum dependent. As long as the beam profile is perfectly flat and the detection efficiency of the proton detector is uniform the two effects cancel. But if the profile is non-flat the probabilities for the two cases become different. This leads to an energy-dependent systematic effect.

The effect can be calculated using Monte Carlo simulations. Still, the neutron apertures were simulated and designed to obtain a beam profile as flat as possible to minimise the effect. To verify our calculations of the edge effect, data was taken for different beam profiles. These profiles were generated with an additional aperture in front of the spectrometer. Each profile was measured by copper foil activations in front of and behind the spectrometer.

\section{Conclusion}
In the recent beam time we showed that our spectrometer is capable of measuring the proton recoil spectrum with high precision. Compared to a previous beam time, the major improvements were the new proton detector, the redesign of several electrodes, and a better vacuum. From the investigated systematic effects and the collected statistics, we expect a total relative error well below 5\,\%.

With the knowledge of the systematic effects gained in this beam time we should be able to improve $a$SPECT to permit a measurement of $a$ with its design accuracy, which is 0.3\,\% in a further beam time.\\

We thank Jim Byrne and Gerd Petzoldt for their contributions in the earlier stages of the experiment, and D. Dubbers and B. M\"arkisch for the $^6$LiF apertures. We also gratefully acknowledge PNSensor GmbH and the MPI Halbleiterlabor for supporting this experiment with their excellent SDDs.

This work was supported by the Bundesministerium f\"ur Bildung und Forschung under Contract No. 06MZ989I, 06MZ170, 06MT196, and 06MT250, and by the European Commission under Contract No. 506065.

\end{document}